\documentclass[preprint,12pt]{aastex}

\usepackage{latexsym}

\shorttitle{Jovian Planets Around White Dwarfs}
\shortauthors{Chu, Dunne, Gruendl, \& Brandner}

\begin{document}

\title{A Search for Jovian Planets around Hot White Dwarfs}

\author{You-Hua Chu, Bryan C. Dunne, Robert A. Gruendl}

\affil{Department of Astronomy,
University of Illinois,
1002 West Green Street, 
Urbana, IL 61801}
\email{chu@astro.uiuc.edu, carolan@astro.uiuc.edu, gruendl@astro.uiuc.edu}

\and

\author{Wolfgang Brandner}
\affil{University of Hawaii,
Institute for Astronomy, 
2680 Woodlawn Drive, 
Honolulu, HI 96822}
\email{brandner@ifa.hawaii.edu}

\begin{abstract}
Current searches for extrasolar planets have concentrated 
on observing the reflex Doppler shift of solar-type stars.  
Little is known, however, about planetary systems around 
non-solar-type stars.  We suggest a new method to extend planetary 
searches to hot white dwarfs.  Near a hot white dwarf, the atmosphere 
of a Jovian planet will be photoionized and emit hydrogen 
recombination lines, which may be detected by high-dispersion 
spectroscopic observations.  
Multi-epoch monitoring can be used to distinguish between non-LTE
stellar emission and planetary emission, and to establish the
orbital parameters of the detected planets.  
In the future, high-precision astrometric measurements of the hot 
white dwarf will allow the masses of the detected planets to be 
determined.  Searches for Jovian planets around
hot white dwarfs will provide invaluable new insight on the 
development of planetary systems around stars more massive than 
the Sun and on how stellar evolution affects these systems. 
We present high-dispersion spectroscopic observations of the 
white dwarf Feige~34 to demonstrate the complexity and feasibility 
of the search method.
\end{abstract}


\newpage

\section{Introduction}

Searches for extrasolar planets have employed four techniques: 
direct imaging of the star and planet, photometric monitoring for
transits, and astrometric and Doppler measurements of the reflex 
motion of the star due to planets.  The imaging technique is 
difficult because a Jupiter-like planet is $\sim$10$^{9}$ and 
$\sim$10$^{4}$ times fainter than a solar-type host star at the 
visible and thermal IR wavelengths, respectively.  The photometric 
method searches for occultations, but can only detect systems
with the orbital plane seen edge-on, which is an exceedingly rare 
circumstance.
For example, a Jupiter-sized planet orbiting a solar-type star at 1~AU
has only $\lesssim$0.5\% chance of being in the proper alignment for
photometric detection.
The reflex motion of a star due to planets is typically small.  To date, 
only the Doppler measurements have been successful in discovering 
planets around a pulsar \citep{Wol92} and solar-type stars 
\citep[e.g.,][]{MQ95,MB96,BM96,CH97}.  
With the exception of the planets around the pulsar 
PSR1257+12, all other known extrasolar planets are Jupiter-sized, 
0.2--11 Jupiter masses (M$_{\rm J}$), with orbital semi-major axis 
ranging from 0.04 AU to 3 AU.  

Current searches for extrasolar planets do not explore whether 
intermediate-mass (2 -- 7 M$_\odot$) stars also possess planetary systems,
because these stars, of MK spectral types A and B, have fewer stellar 
absorption lines available for accurate velocity measurement.  Furthermore,
because they are more massive than solar-type stars, the stellar
reflex motion is even smaller.  These difficulties can be circumvented, 
if one searches for planets around the descendants of these 
intermediate-mass stars: hot white dwarfs with $T_{eff} > 50,000$ K.  
A hot white dwarf's UV radiation will photoionize the planetary 
atmosphere and recombination lines will be emitted by the planet.  
The contrast between the planet's recombination line emission and the 
stellar spectrum will be the limiting factor for detecting 
the planet.  The strength of the planet's recombination line emission 
is proportional to the UV flux where the stellar emission
peaks, but the recombination line is emitted at a longer wavelength 
where the stellar emission is lower; therefore, the contrast between 
the planet's recombination line emission and the stellar spectrum
is not as extreme as the contrast between the planet and star 
both in continuum.  

Thus, it may be possible to find planets around hot white dwarfs by
searching for a periodically Doppler-shifted recombination line 
emission superposed on the stellar spectrum.  We have carried out 
calculations to demonstrate the feasibility of this search, taking 
into account the stellar evolutionary effects.  In this letter we 
report the results of our calculations, discuss practical 
considerations, and present observations of a white dwarf to 
illustrate observational considerations.

\section{Detectability of Planets around Hot White Dwarfs}

Our proposed method to search for planets around hot white dwarfs
utilizes recombination lines emitted by the ionized atmosphere of 
orbiting planets.  In order to detect the planetary emission, the 
planet's atmosphere needs to contain an abundant element that has 
recombination lines in easily observable wavelength ranges.
Given currently available facilities, hydrogen provides
the most observable recombination lines.  Therefore, we will consider 
searching for only Jupiter-like gas giants around hot white dwarfs.

The progenitor of a hot white dwarf with $T_{eff} > 50,000$ K has
gone through various evolutionary stages, among which the asymptotic 
giant branch (AGB) phase is the stage at which the star has the most 
pronounced effects on its planetary system, because the star swells up 
to a few hundred R$_\odot$, and goes through episodes of copious mass 
loss \citep{Iben83}.  At least four competing factors may affect 
the planetary system: (1) Planets engulfed in the stellar envelope will
experience a drag force and migrate inward.  (2) The decreasing
central star mass will cause the planets to migrate outward.
(3) Tidal interaction between a planet and the stellar envelope
will slow down the planet's orbital motion and cause it to 
migrate inward. (4) Planets orbiting an AGB star can experience
accretion and/or evaporation.  These factors have been taken into
account by \citet{LS84} in their calculation of the evolution of a 
star-planet system.  They found that Jovian planets engulfed in the
AGB envelope will either evaporate or accrete mass to become a 
close companion, depending on the mass of the stellar envelope,
the initial mass of the planet, and the initial separation between
the planet and the star.  The calculations of \citet{LS84} are
admittedly qualitative, and there are no theoretical predictions
which preclude the existence of Jovian planets within a few AU 
of a post-AGB star, or a hot white dwarf.  Furthermore, a Jovian
planet has been suggested to be a companion to the hot white
dwarf PG~0950$+$139 \citep{DL89}.

Hot white dwarfs with $T_{eff} > 50,000$ K are powerful sources for 
H-ionizing UV radiation.  In this environment, the hydrogen-rich outer 
envelopes of any Jovian planets present will be photoionized.  The 
resulting recombinations will cause the planet to emit hydrogen 
recombination lines.  To calculate the recombination line strengths 
we have made a number of assumptions and approximations.  For the
stellar radiation, we use a blackbody radiation model to approximate
the stellar spectrum and use 1 R$_\oplus$ as the stellar radius
to determine the ionizing photon flux emitted by the star.
We approximate the planetary atmosphere as a plane-parallel slab
with physical dimension equivalent to the cross section of the 
planet.
Within the slab, the UV flux from the white dwarf
will ionize the outer layer of the atmosphere, to the point where
the  number of incident ionizing photons balances the number 
of recombinations.
In other words, we treat the atmosphere as a ``Str\"omgren slab."
Assuming the ionized atmosphere is at a temperature of 10,000 K, the 
expected H$\alpha$ luminosity is then given by:

\begin{equation}
L_{{\rm H}\alpha} = 7.7\times 10^{-20}~Q~({{r_{\rm P}}\over{a}})^2~
~~~~{\rm ergs~s^{-1}},
\label{eqn:lum}
\end{equation}

\noindent where $Q$ is the ionizing flux of the white dwarf in number of 
ionizing photons per second, $r_{\rm P}$ is the radius of the planet
in Jupiter radii (R$_{\rm J}$), and $a$ is the planet's orbital 
semi-major axis in AU.

\begin{figure}[th]
\epsscale{0.75}
\centerline{\plotone{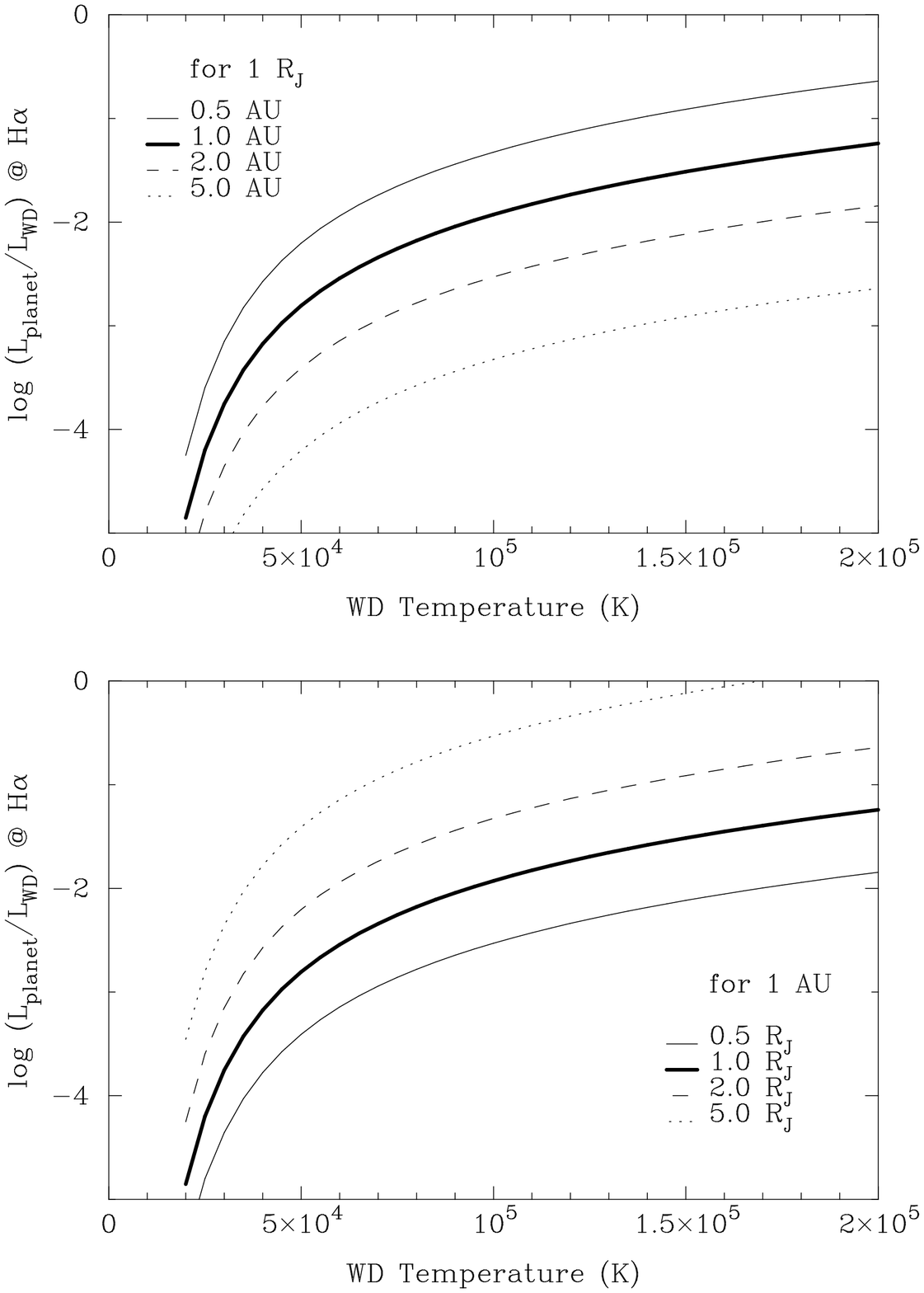}}
\figcaption[Chu.fig1.eps]{Prediction of the contrast between 
the H$\alpha$ line emission from a planet (L$_{\rm planet}$) and 
the continuum emission from the host white dwarf (L$_{\rm WD}$) 
as a function of $T_{eff}$ of the white dwarf.  The upper panel shows 
the variation in contrast for a 1 R$_{\rm J}$ sized Jovian planet
at four different orbital semi-major axes.  
The lower panel shows the variation in contrast 
with respect to the planetary radius for an orbit with semi-major axis 
of 1 AU.  See text for uncertainties in the model. \label{fig:slab}}
\end{figure}

We have estimated the expected brightness contrast between
a Jovian planet and the host star at the H$\alpha$
line using the model described above.  The calculations were
performed for planets with radii of 0.5, 1.0, 2.0, and 5.0 R$_{\rm J}$ 
and orbital semi-major axes of 0.5, 1.0, 2.0, and 5.0 AU
over a range of white dwarf temperatures, 20,000--200,000 K.   
The ratio of the Jovian planet's H$\alpha$ line flux to the 
underlying stellar flux (within a comparable bandwidth, 
$\sim0.5$ \AA, the expected thermal width of the H$\alpha$ line) 
is shown in Figure~\ref{fig:slab}.  

From this simplistic model, it can be seen in Figure~\ref{fig:slab}
that the ratio between the planet's H$\alpha$ line emission and the 
stellar continuum emission can be as high as $\sim$1/20 for a 
Jupiter-sized planet at 1 AU from a $\sim$200,000 K star.
This model, however, ignores many effects which will lower the 
contrast between the planet and star.  For instance, in a dense,
molecule-rich planetary atmosphere, H$^+$ can react with atoms
or molecules rather than recombine with a free electron; thus,
the planetary H$\alpha$ emission will be reduced.
Furthermore, many hot white dwarfs show narrow H$\alpha$ and
\ion{He}{2} $\lambda$6560 line emission superposed on broad H$\alpha$ 
and \ion{He}{2} absorption \citep{RW89} because of non-LTE 
(non-Local Thermal Equilibrium) effects in the 
stellar atmosphere \citep{LH95,Hubeny99}.  This will complicate
the search for planetary H$\alpha$ emission.
Therefore, the results in Figure~\ref{fig:slab} portray an overly
optimistic view of the detectability of Jovian planets around a hot
white dwarf.  On the other hand, the photoionization-driven ablation
of a planetary atmosphere may produce an extended low-density 
($\sim10^6$ H-atom~cm$^{-3}$) nebula, and the forbidden lines
emitted form this nebula can be easily detected \citep{DL89}.

\section{Observational Considerations}

To detect the H$\alpha$ emission from a Jovian planet around a hot
white dwarf, multi-epoch, high-dispersion, high-S/N spectroscopic 
observations of a large sample of hot white dwarfs are needed.
Below are some of the observational considerations for such a search.

\begin{itemize}
\item Velocity resolution: The orbital velocity of a planet within 
5 AU of the host star will be $\gg$10 km~s$^{-1}$.  To be able to 
detect this motion, the precision of the velocity measurements should be  
$\sim$1--2 km~s$^{-1}$.  This can be easily achieved with a 
high-dispersion spectrograph with R = 30,000.

\item Large telescope: A sensitive search for recombination line emission
from a planet requires a high signal to noise ratio (S/N).  To achieve 
a S/N ratio of 100 for a 13.5 mag star with a reasonable exposure time
($\lesssim$1.5 hours), the telescope must be at least $\gtrsim$4~m in
diameter. 

\item Long slit: Due to the possible faintness of the planetary 
recombination line emission, telluric and interstellar line emission
can interfere with the detection of the planetary lines.
To facilitate the removal of telluric and interstellar line emission,
a long slit is required for the observations.

\item Multi-epoch observations: Given the range of semi-major axes
over which this method is likely to detect H$\alpha$ emission from a 
planet (a few AU), multiple epoch observations separated by 4--6 months
are required to detect any change in the H$\alpha$ emission line due to 
the orbital motion of the planet.  Planetary H$\alpha$ emission may
be distinguished from stellar H$\alpha$+\ion{He}{2} emission by the 
periodic velocity change of the planetary emission.
Monitoring observations over a period
of 2--3 years will be needed to determine the planetary orbital elements.

\item Planetary mass measurements: Unlike reflex motion planet searches, 
the velocity shift observed is that of the planet rather than that 
of the host star.  The mass of the planet cannot be accurately determined
from this motion.  Therefore, astrometric measurements of the reflex motions
of the host white dwarfs are needed to determine the masses of discovered 
planets.  

\end{itemize}

\section{Example Observation of a Hot White Dwarf}  

\begin{figure}[th]
\epsscale{0.9}
\centerline{\plotone{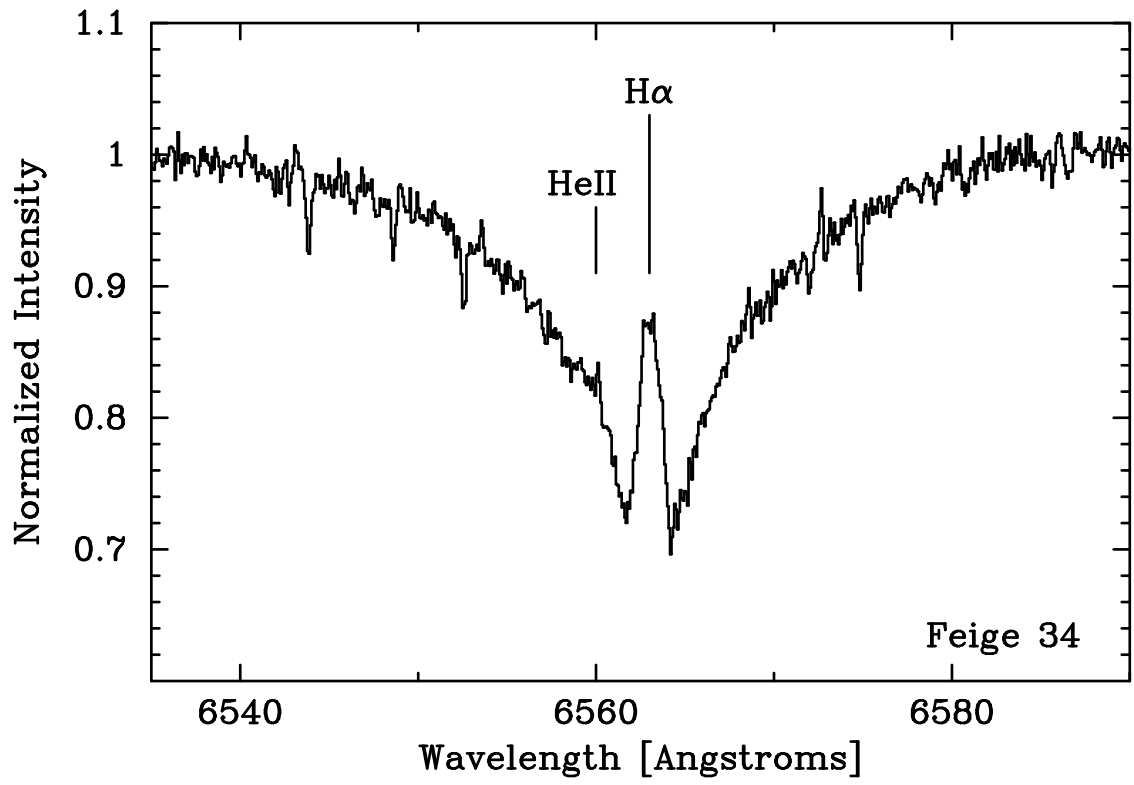}}
\figcaption[Chu.fig2.eps]{Continuum-normalized H$\alpha$+\ion{He}{2} line
profile of Feige 34. \label{fig:echelle}}
\end{figure}

The hot white dwarf Feige~34 was observed on 2000 April 23 (UT) with 
the echelle spectrograph on the 4-m telescope at Kitt Peak National 
Observatory.  The high-dispersion spectra were obtained
using the \mbox{79 l mm$^{-1}$} echelle grating in combination with
a \mbox{226 l mm$^{-1}$} cross disperser and the long focus red camera.
A reciprocal dispersion of \mbox{3.5 \AA\ mm$^{-1}$} at H$\alpha$ 
was achieved with this setup.  The spectra were imaged with the 
T2KB CCD detector.  The pixel size is 24~$\mu$m, corresponding 
to 0\farcs24~pixel$^{-1}$ along
the slit and \mbox{$\sim$3.7 km s$^{-1}$ pixel$^{-1}$} along the 
dispersion axis.  Two 1800~s exposures were taken of Feige~34 with 
an east--west oriented 20\arcsec-long slit of width 1\farcs5.  
The H$\alpha$ region of the echellogram is presented in 
Figure~\ref{fig:echelle}.  All of the data were reduced using 
standard routines in IRAF\footnote{Image Analysis and Reduction
Facility -- IRAF is distributed by the 
National Optical Astronomy Observatories, which are operated by 
the Association of Universities for Research in Astronomy, Inc., 
under cooperative agreement with the National Science Foundation.}.

The echellogram of Feige~34 shows broad H$\alpha$+\ion{He}{2} absorption  
from the white dwarf superposed with a narrow H$\alpha$ emission component
and a faint \ion{He}{2} emission component (marked in 
Figure~\ref{fig:echelle}).
The H$\alpha$ emission component has a FWHM of $\sim$70 km~s$^{-1}$;
it appears offset blueward from the center of the 
absorption because the absorption is a blend of a strong H$\alpha$ 
line and a weak \ion{He}{2} line.
Using the observations of \citet{Massey88} for a flux 
calibration, we determine the H$\alpha$ flux of the emission component to be
\mbox{$1.4 \times 10^{-14}$ erg cm$^{-2}$ s$^{-1}$}.  This large
flux and the lack of interstellar emission features on the 
Palomar Sky Survey plates rules out an interstellar origin for this
emission component.  Furthermore, the lack of forbidden line emission
excludes the possibility that the H$\alpha$ emission arises from a 
typical circumstellar nebula.  It is possible that this H$\alpha$
emission is dominated by the stellar H$\alpha$ emission due to a 
non-LTE atmosphere \citep{Werner2000}.
This does not, however, exclude the possibility that the non-LTE 
stellar emission is superposed with weak H$\alpha$ emission from an 
orbiting planet or low-mass dwarf companion.

A low-mass red dwarf companion to Feige~34 has been suggested by
\citet{Thejll91} based on an infrared excess in JHK bands.  
Alternatively, a large planet close to the star, heated by the 
UV flux, may also produce the observed infrared excess.  
To distinguish between these two possibilities, further theoretical 
and observational work is required.
Detailed modeling of a non-LTE white dwarf atmosphere for Feige~34 
is needed to quantitatively assess whether all or part of the H$\alpha$ 
emission feature in the spectrum of Feige~34 arises from the star.  
Follow-up spectroscopic observations are necessary to determine
whether the H$\alpha$ emission line profile shows temporal variations
suggesting the existence of a companion.  If temporal variations are
detected, further monitoring observations would allow the orbital 
parameters and nature of the companion to Feige~34 to be determined.

\section{Summary}

We suggest that high-dispersion spectroscopic observations of hot white 
dwarfs can be used to search for H$\alpha$ emission from Jovian planets 
within a few AU of the star.  H$\alpha$ emission from the ionized outer
envelope of a Jovian planet as weak as 1/100 the stellar emission
can be detected with modern large ($\gtrsim$4-m) telescopes.
Non-LTE effects in the hot white dwarf atmosphere may produce stellar 
H$\alpha$ emission.  Stellar and planetary H$\alpha$ emission can be
disentangled by multi-epoch observations if changes in the H$\alpha$ 
emission line profile due to the orbital motion of the planet are 
detected.  Monitoring observations to establish the orbital parameters
are required to distinguish between Jovian planets and a low-mass 
dwarf companion.  
High-dispersion spectroscopic observations of Feige~34 demonstrate
the complexity and feasibility of this method to search for these
planets.  Such a search will provide targets for future 
high-precision astrometric observations (e.g., SIM, adaptive optics,
etc.), and valuable insight can be obtained on the development and 
evolution of planetary systems around stars more massive than the Sun.

\acknowledgements{We would like to thank Klaus Werner for enlightening
discussions on non-LTE effects in a hot white dwarf atmosphere and
sharing his model calculations of Feige~34's H$\alpha$+\ion{He}{2}
lines.  We also thank Ron Webbink, Charles Gammie, and Wing Ip for 
helpful conversations,  the anonymous referee for insightful 
comments, and Caty Pilachowski for sharing unpublished high-dispersion 
spectra of white dwarfs.}

\clearpage

\end{document}